\documentclass[sigconf]{acmart}

\usepackage{subfigure}
\usepackage{tabularx}
\usepackage{dcolumn}
\usepackage{algorithm,algpseudocode}
\algnewcommand\algorithmicinput{\textbf{Input:}}
\algnewcommand\algorithmicoutput{\textbf{Output:}}
\algnewcommand\Input{\item[\algorithmicinput]}%
\algnewcommand\Output{\item[\algorithmicoutput]}%

\usepackage{caption}
\newcolumntype{d}[1]{D{.}{.}{#1}}

\AtBeginDocument{%
  \providecommand\BibTeX{{%
    \normalfont B\kern-0.5em{\scshape i\kern-0.25em b}\kern-0.8em\TeX}}}

\setcopyright{acmcopyright}
\copyrightyear{2018}
\acmYear{2018}
\acmDOI{10.1145/1122445.1122456}

\acmConference[Open research]{}{January 2021}{Remote}
\acmBooktitle{Open research}
\acmPrice{0.00}
\acmISBN{}

\acmSubmissionID{1716}

\begin{document}

\title{Lightweight representation learning for efficient and scalable recommendation}


\author{Olivier Koch}
\email{o.koch@criteo.com}
\affiliation{%
  \institution{Criteo}
  \country{France}
}

\author{Amine Benhalloum}
\email{ma.benhalloum@criteo.com}
\affiliation{%
  \institution{Criteo}
  \country{France}
}

\author{Guillaume Genthial}
\email{g.genthial@criteo.com}
\affiliation{%
  \institution{Criteo}
  \country{France}
}

\author{Denis Kuzin}
\email{d.kuzin@criteo.com}
\affiliation{%
  \institution{Criteo}
  \country{France}
}
  
\author{Dmitry Parfenchik}
\email{d.parfenchik@criteo.com}
\affiliation{%
  \institution{Criteo}
  \country{France}
}

\renewcommand{\shortauthors}{}

\begin{abstract}

Over the past decades, recommendation has become a critical component of many online services such as media streaming and e-commerce. Billion-scale recommendation engines are becoming common place. Meanwhile, new state-of-the-art methods such as variational auto-encoders have appeared, raising the bar in performance. 

We propose to go a step further with a simple and efficient model (LED, for Lightweight Encoder-Decoder). By combining pre-training, sampled losses and amortized inference, LED brings a $30\times$ inference speed-up compared to the best system known so far, while reaching the performance of variational auto-encoders on standard recommendation metrics.

LED has been deployed in production at Criteo and serves billions of users across hundreds of millions of items in a few milliseconds using standard hardware. We provide a detailed description of our system and illustrate its operation over two months of experiment. We also release the code for LED. Our work should be of significant interest to practitioners wishing to deploy an efficient large-scale recommendation system in the real-world.
\end{abstract}

\ccsdesc[500]{Information systems~Display advertising}
\ccsdesc[500]{Computer systems organization~Real-time system architecture}
\ccsdesc[300]{Computing methodologies~Machine learning algorithms}

\graphicspath{ {fig/} }

\keywords{online advertising, recommender systems, representation learning}


\maketitle

\section{Introduction}
\label{introduction}

Online advertising offers a unique test bed for recommendation at scale. Every day, billions of users interact with millions of products on thousands of retailers' websites in real-time. Systems designed to address this challenge must provide a recommendation for any user in a few milliseconds while leveraging the massive amounts of data available for training. 


Building a successful recommendation engine in this context requires a scalable and deep understanding of the users, the merchants, the products, and their respective interactions. Representation learning is an attractive approach as it provides a way to embed various entities into the same space which can be then searched through nearest-neighbor efficiently. Recent years have seen tremendous progress in the field, giving birth to a large variety of approaches such as variational auto-encoders~\cite{variational-liang-2018, gated-vaes, rec-vae, ease-vae}, reinforcement learning~\cite{ract-cf} and graph networks~\cite{graph-networks-pinterest-kdd-2018, alibaba-intentgc-kdd-2019, alibaba-m2glr-kdd-2020, pixie-pinterest-www-2018}, to name a few. Meanwhile, billion-scale recommender system have become common place~\cite{pixie-pinterest-www-2018, alibaba-intentgc-kdd-2019, alibaba-m2glr-kdd-2020, alibaba-graphs-kdd-2018}. 

Making these systems efficient is critical in order to make state-of-the-art methods accessible to many in the real world and reduce their computational footprint. In addition, the end-to-end problem of large-scale training and real-time inference is rarely addressed. 

Our contribution is to make a new leap forward in terms of simplicity and computing efficiency while maintaining state-of-the-art performance at scale. We analyze the fundamental bottlenecks of existing state-of-the-art algorithms. We then devise several practical design choices which, once applied to these algorithms, allows us to design a new method (LED, for Lightweight Encoder-Decoder). Our experiments show that LED brings a $30\times$ speed-up in latency against the best well-known baseline~\cite{pixie-pinterest-www-2018} while reaching the algorithmic performance of state-of-the-art methods~\cite{variational-liang-2018}.

LED has been deployed at Criteo and powers our recommendation engine in production. We provide a detailed description of our system architecture, from offline training to online inference, delivering $3200$ queries per second with sub-millisecond latency. We further demonstrate its operation over several months at scale and open-source the code. Our work should  be of interest to practitioners wishing to deploy an efficient large-scale recommendation system in the real-world.

\section{Problem statement}

\subsection{Definitions}

We consider the general problem of recommending products to users on the internet. Products are shown on banners displayed on publisher websites at the scale of billions of users, dozens of millions of items, and billions of displays per day. 


A product is any item a user can purchase: retail, travel, etc. We assume that events coming from the same user across merchants and banners are aggregated into a single series of events called a {\it user timeline}. 
We also assume that an attribution mechanism associates a product sale on a merchant to the last product clicked on a previously displayed banner (if such an event ever occurred).

The goal of the recommendation engine is to show products that will be of interest to the user, captured either through clicks on the banners or through sales on the merchants' websites. 
 






\subsection{Requirements}
\label{subsec:requirements}

Our system needs to address the following requirements:

\textbf{Scale} The recommendation system must work at the scale of billions of users and hundreds of millions of items.


\textbf{Latency} The system should respond within a few milliseconds to fit the need for banner display on mobile and web applications. 


\textbf{Churn} Users enter and exit the system at a far higher rate than changes in the product catalog. The system should avoid having to recompute its parameters for each update in the user base.



\textbf{Multiple feedbacks} Collected data usually involves different types of feedback (product views and sales, clicks on banners). We seek a design that will extract as much information as possible from the available data.



\section{Related Work}
\label{related-work}

\textbf{Representation learning for recommendation} A classical approach to recommendation uses collaborative filtering through factorization of the user-item matrix~\cite{wmf}. This approach leverages decades of research and produces robust results at scale ~\cite{amazon,item-based-collaborative-filtering}. The field went through a significant renewal when randomized algorithms helped scale to large dimensions~\cite{finding-structure-2011, constantine-2011}. SLIM~\cite{slim} and its variants~\cite{efficient-slim,ease-vae} differ from standard matrix factorization by solving a constrained linear problem to learn a sparse item-item matrix. Recent improvements like EASE~\cite{ease-vae} relax some of the constraints to find a dense closed-form solution. These methods are slow to train at a large scale~\cite{variational-liang-2018}, even if quantization and fast nearest-neighbor techniques can speed up inference.

A vast body of literature recently grew around neural networks, leveraging various architectures such as convolutional, recurrent and graph networks~\cite{deep-learning-karatzoglou-2017, deep-music-recommendation-nips, deep-survey, graph-networks-pinterest-kdd-2018, alibaba-intentgc-kdd-2019, alibaba-m2glr-kdd-2020, pixie-pinterest-www-2018, alibaba-ctr}. These models allow to seamlessly fit into a single loss a number of constraints or metadata available in the input. The affinity between a user $ u $ and an item $ i $ is usually modelled as a score function $ \hat{y}(u, i; \theta) $, where $ \hat{y} $ is a neural network with parameters $\theta $. While most methods express the final score via inner product~\cite{lightrec-www-2020,gramian}, or euclidean distance~\cite{cml-www}, making it possible to use \emph{fast-KNN} retrieval techniques, some~\cite{ncf-www-2017} require to rank all products, making inference cost linear in the number of products. 

More recently, Variational Auto-Encoders~\cite{variational-liang-2018,rec-vae,gated-vaes} generalizing latent factor models have established a new standard. While integrating new users is made possible by using bag-of-words representations, the multinomial likelihood commonly used as objective involves a softmax over all items, which becomes too computationally expensive at the scale at which we operate. Improvements~\cite{ract-cf} using reinforcement learning have further increased their performance, but lack the efficiency needed at large scale.

\textbf{Billion-scale recommendation systems}. With the advent of ever newer recommendation algorithms, a major concern of the community has been to bring them to the scale of the internet. Early work demonstrates a deep learning architecture operating at the scale of Youtube~\cite{deep-youtube}. Compact hierarchical networks~\cite{aws-hrnn-kdd-2020} address real-time and diverse metadata needs and have been deployed at Amazon Web Services. Neural input search~\cite{google-neural-search-kdd-2020} learns vocabulary and embedding sizes and allows the embedding dimension to vary for different values of the feature. Recent work investigates how
to correct biases induced by two-stage recommender systems with off-policy reinforcement learning at scale~\cite{off-policy-reco-www-2020}.  



Advances at scale are particularly notable on graph networks. A data-efficient Graph Convolutional Network (GCN) combines efficient random walks and graph convolutions incorporating both graph structure and node information at the scale of billions of users and items~\cite{graph-networks-pinterest-kdd-2018}. An efficient graph convolutional model~\cite{alibaba-intentgc-kdd-2019} captures both explicit user preferences and heterogeneous relationships of side information over hundreds of millions of users and items. Multi-task multi-view graph representation learning fuses multiple representations for 57 billion examples~\cite{alibaba-m2glr-kdd-2020}.  Finally, graph embeddings with side information~\cite{alibaba-graphs-kdd-2018} leverage unsupervised feature learning at the scale of billions of users and billions of items. 

Latency and load are rarely described in the literature, yet represent major constraints in real-world applications. The best performance we are aware of is reached by Pixie at Pinterest, leveraging graph networks to handle $1,200$ recommendation queries per second (qps) with 60 millisecond latency~\cite{pixie-pinterest-www-2018}. Collaborative Metric Learning~\cite{cml-www} leverages LSH to achieve $10,000$ qps on a catalog of $260,000 $ items. Our contribution is to show that a lightweight encoder-decoder model (LED) reaches the performance of VAEs while allowing a $30\times$ speed-up in latency and high throughput.


\section{Efficient models}
\label{sec:models}

We present here the main design choices enabling efficient representation learning at scale. We use $i \in \left\{ 1, ..., I \right\} $ to index items, $u \in \left\{ 1, ..., U \right\} $ to index users, and  $u_{(t)} = (u_1, ..., u_T) $ to index the sequence of items one user interacted with. We associate each item $ i $ and user $ u $ with their $ d $-dimensional vector representations $\overrightarrow{v_i} \in \mathbb{R}^d $ and $\overrightarrow{u} \in \mathbb{R}^d $.


\textbf{Fast nearest-neighbor search} For a given user $ u $, the system should rank items with a scoring function expressed as an inner product $ s(u, i) = \langle \vec{u}, \vec{v_i} \rangle $. Finding the $ k $ best items for user $ u $ is thus equivalent to finding the $ k $ nearest neighbors of $ \vec{u} $ with maximum inner product. Thanks to this formulation, we can leverage efficient approximate nearest neighbor techniques (see Section \ref{subsec:online-knn}) at inference.

\textbf{Amortized inference} The use of amortized inference~\cite{amortized-inference} consists in sharing the same procedure to compute user representations, effectively making the number of parameters to learn independent from the number of users and addressing user \emph{churn}. This is one of the strengths of the VAE framework~\cite{ract-cf}.

\textbf{Sampling-based losses} The computational complexity of comprehensive losses like the multinomial and gaussian likelihoods is linear in the number of items which makes them unusable under our requirements. Instead, we can use \emph{ranking}-based losses, such as Bayesian Personalized Ranking (BPR)~\cite{bpr}, Negative Sampling (NS)~\cite{NIPS2013_5021}, or \emph{approximation}-based methods like Complementarity Sum Sampling (CSS)~\cite{pmlr-v54-botev17a}. 

\textbf{Pre-training} While the final recommendation system is evaluated on \emph{click} events, we can leverage the large amount of historical data (mainly \emph{view} events) to pre-train the model. In practice, we pre-train item embeddings $ \vec{v_i} $ on \emph{view} events using large-scale matrix factorization (see Algorithm~\ref{alg:led-training-procedure}). 

\subsection{Pre-training with large-scale matrix factorization Randomized SVD}
\label{subsec:matrix-fatorization-with-randomized-svd}

We first present a \emph{pre-training} method to get item embeddings consisting in factorizing an item-item matrix. We depart from the usual user-item rating matrix factorization since having latent representations for users would not be robust to churn. Also, the number of users is the limiting factor regarding scalability as it greatly exceeds the number of items.

Instead, we factorize the Pointwise Mutual Information (PMI~\cite{pmi-bouma-2009}) matrix.  Writing $ p(i) $ the probability of item $ i $ to appear in the data, $ p(i, j) $ the probability of items $ i $ and $ j $ to co-occur in one user timeline, the PMI coefficient associated with items $ i $ and $ j $ is defined as $ \text{PMI}_{i,j} = \log (p(i, j)/\left(p(i)p(j) \right))$. This allows us to capture similarities between all items, since the coefficients are normalized by popularity. The number of views being unevenly distributed, factorizing the co-occurrence matrix would give too much importance to popular items. 





Since we collect cross-merchant interactions, the resulting matrix cannot be factorized by block and we need to leverage factorization methods that can operate on sparse matrices with hundreds of millions of rows and columns. We use Randomized Singular Value Decomposition (RSVD). RSVD first projects the input matrix into a low-dimensional space which captures most of the norm of the input matrix. Second, using the QR-decomposition of this thin matrix~\cite{constantine-2011} it computes its left singular vectors which we use as item embeddings. 



\subsection{Lightweight Encoder-Decoder (LED)}
\label{subsec:LED}

We derive a simple and efficient model, the \emph{Lightweight Encoder-Decoder (LED)}, from a common yet powerful baseline~\cite{starspace-facebook}. This model has few parameters, one embedding $ v_i \in \mathbb{R}^d $ and bias $ b_i \in \mathbb{R} $ per item. It encodes each user timeline $ u_1, ... u_T $ to derive the user representation $ \vec{u} $ via a simple average, respecting the \emph{amortized inference} guideline:
\begin{align*}
    \overrightarrow{u} &= \frac{1}{T}\sum_{t=1}^{T} \overrightarrow{v_{u_t}}.
\end{align*}
It then "decodes" the user representation to retrieve recommendation scores for each item $ i $ as follows:
$$
s(u,i) = \langle  \overrightarrow{u},  \overrightarrow{v_i} \rangle + b_i.
$$

Because some items tend to be more popular than others, we capture the global tendency with biases $ b_i $. As a result, the inner product can focus on modeling the true interaction between a user and an item. While using biases as a first-order approximation of the user-item matrix is standard in the matrix factorization literature~\cite{koren,paterek}, we differ from it by jointly training embeddings and biases instead of using a pre-computed average.

Note that the score can be rewritten as a plain inner product if we add an extra dimension to the vectors $ \vec{u} $ and $ \overrightarrow{v_i} $ with values $ 1 $ for the user and $ b_i $ for the item, making it compatible with our \emph{fast nearest-neighbor search} guideline.

\begin{algorithm}[tb]
  \begin{algorithmic}[1]
    \Input User Timelines (views and clicks on items)
    \Output Item embeddings $ v_i = P \cdot \overrightarrow{v_i^{\text{RSVD}}} \in \mathbb{R}^d $ and biases $ b_i $
    \State Build the PMI matrix $ \in \mathbb{R}^{I \times I} $ on \textbf{view} events
    \State Run RSVD on the $ \text{PMI} $ matrix to get pre-trained item embeddings $ \overrightarrow{v_i^{\text{RSVD}}} \in \mathbb{R}^d $
    \State Randomly initialize parameters $ P \in \mathbb{R}^{d \times d} $ (projection matrix) and $ b_i \in \mathbb{R} $ (items' biases). Given a user $ u $ and item $ i $, define
    $$ s(u, i) = \langle \vec{u} , P \cdot \overrightarrow{v_i^{\text{RSVD}}} \rangle  + b_i \text{ \ with \ } \vec{u} = \frac{1}{T} \sum P \cdot \overrightarrow{v_{u_t}^{\text{RSVD}}} $$
    \State For each user $ u $ associated with a positive \textbf{click} event $ p $, sample a negative item $ n $ and update $ P $ and $ b_i $ to maximize the BPR
        $$ \log \sigma \left( s(u, p) - s(u, n) \right)$$
    
\end{algorithmic}
 \caption{LED: Training with \emph{project} fine-tuning}
 \label{alg:led-training-procedure}
\end{algorithm}

\subsection{Sampling-based losses}
\label{subsec:training}



We split the timelines into two parts, \emph{input} and \emph{target}. The model is trained to predict the content of the \emph{target} period given the \emph{input} period. In the case where we have two types of events (\emph{view} and \emph{click}), we use all events as inputs and only include \emph{click} events in the target. Similarly to the Mult-DAE, we randomly drop a fraction of the \emph{input} items. It reduces over-fitting and leads to competitive results, as shown in ~\cite{variational-liang-2018}.

In all generality, we want $ s(u, i) $ to be high for the items in the \emph{target} period. We consider the following loss functions.





\textbf{Approximated Mult-Likelihood} A common approach consists in treating the recommendation problem as a multi-class multi-label classification problem, where the labels are the items in the \emph{target}. Traditionally used in the Auto-Encoder literature, the gaussian likelihood has been shown to be less effective than the Multinomial Likelihood~\cite{variational-liang-2018}, which models each item's probability using a softmax such that $ \pi(i | u) \propto \exp{(s(u, i))}$ and assumes that the user's items are drawn from $ \pi $ according to a multinomial distribution. Unfortunately, using a softmax involves a partition function $ Z(u) = \sum_{i=1}^I \exp{(s(u, i))} $ whose computation is too costly.



We circumvent this problem using Complementarity Sum Sampling (\emph{CSS}) \cite{pmlr-v54-botev17a} which provides a robust approach to estimate $ Z(u) $ via sampling. In practice, we sample $ N $ negatives for each target item $ i $ and estimate its partition function as:
$$
\widehat{Z_i} = \exp{(s(u, i))} + \frac{I - 1}{N} \sum_{n=1}^{N} \exp(s(u, n)).
$$

\textbf{Ranking} Another approach is to see the task as a ranking problem. In other words, the objective is to differentiate the \emph{positive} item from sampled \emph{negative} items. In our settings, we use the Bayesian Personalized Ranking (BPR) loss \cite{rendle2009bpr}, which maximizes the ranking
$$
\sum_{n=1}^N \log \sigma(s(u, i) - s(u, n))
$$
where $ \sigma $ is the sigmoid function. We note that the CSS method with one negative is equivalent to a BPR loss with margin $ \log(I - 1) $.

\textbf{Negative Sampling} Traditionally used in Natural Language Processing ~\cite{NIPS2013_5021} and similar in its form to the ranking formulation, it consists in training the model to distinguish positives from negatives by maximizing
$$
\log \sigma(s(u, i)) + \sum_{n=1}^N \log \left( 1 - \sigma(s(u, n)) \right).
$$

\subsection{Fine-tuning RSVD Embeddings}
\label{subsec:fine-tuning-svd-embeddings}

Item embeddings can be initialized with the RSVD embeddings. Since these embeddings are trained on \emph{view} events, they would benefit from additional training on \emph{click} events. The classical approach is to train them alongside the other model's parameters. We refer to this fine-tuning method as \emph{full}. 

\begin{figure}[t]
\centering
\includegraphics[width=0.30\textwidth]{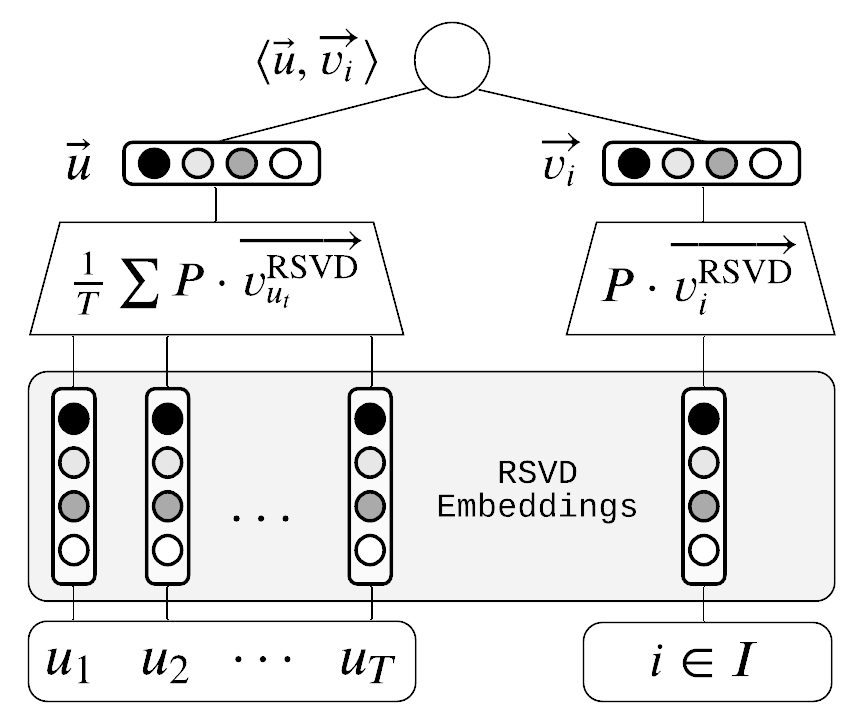}
\caption{LED architecture. P is a trainable projection matrix~$\in~\mathbb{R}^{d \times d}$. Popularity bias omitted for simplicity.}
\label{fig:diagram-learning-led}
\end{figure}

We consider a variant called \emph{project} (see Figure~\ref{fig:diagram-learning-led}), which consists in computing the item embeddings from the RSVD embeddings $ v_i^{\text{RSVD}} $, freezed during training. More specifically, the model learns a projection matrix $ P \in \mathbb{R}^{d \times d} $ such that
$$ 
\vec{v_i} = P \cdot \overrightarrow{v_i^{\text{RSVD}}}.
$$

Using such a simple transformation allows us to rewrite the scoring function using a \emph{transposition trick}:


$$ s(u, i) = \langle P^T \cdot P \cdot \overrightarrow{u^{\text{RSVD}}}, \overrightarrow{v_i^{\text{RSVD}}} \rangle + b_i $$. 

Because we have multiple positives and negatives for a given user, we reduce the number of matrix multiplications. This yields a significant speedup during training (x3 in our setup).

The \emph{project} fine-tuning method is a key factor in the performance of our method. It can be applied to other models (like VAEs), though the transposition trick is only possible in our formulation. By reducing the number of trainable parameters, it speeds up training and decreases exposure to over-fitting. At retrieval time, the computational cost of the \emph{project} method is identical to the \emph{full} method, since we can directly use the projected embeddings.

\subsection{Complexity analysis}

The number of parameters of \emph{LED} is linear in the number of items ($ I \times d $ item embeddings and $ I $ biases), but independent from the number of users $ U $ (contrary to NCF\cite{ncf-www-2017} or matrix factorization methods~\cite{wmf}). When using the \emph{Project} fine-tuning method, the number of trainable parameters is even smaller, $ d \times d + I $.

Training requires multiple updates on each user timeline. Each update using one of the sampled losses is linear in the number of items in the timeline and the number of negatives $ N $. As a result, the final training complexity is $ \mathcal{O}\left(U \times T \times N \right) $, which has the advantage of not being quadratic in $ U $ or $ I $ (unlike EASE~\cite{ease-vae} in $ \mathcal{O}(I^{2.376}) $ or the \emph{Mult-VAE}~\cite{variational-liang-2018} in $ \mathcal{O}(U \times I) $). 


At inference, computing the user representation $ \overrightarrow{u} $ is only linear in the number of items in the input timeline while finding the top-$k$ best recommendations is roughly $ \mathcal{O}(\log(I)) $ thanks to efficient approximate nearest neighbors techniques.

\section{Offline evaluation}
\label{sec:offline-evaluation}
We now evaluate the impact of model choices described in Section~\ref{sec:models}. We insist on the fact that our goal is not to outperform state-of-the-art methods on performance but to reach state-of-the-art performance while operating at much better latency and scale. We compare the performance of \emph{LED} to \emph{Mult-VAE}~\cite{variational-liang-2018} on two datasets using two standard metrics, recall@k and click rank. In particular, we assess the impact of sampling-based losses as well as pre-training. The findings of this section can be summarized as follows:

\begin{itemize}
    \item Training models using a sampling-based loss (\emph{Mult-CSS}\cite{pmlr-v54-botev17a}, \emph{BPR}\cite{bpr} or \emph{NS}\cite{NIPS2013_5021}) gives results close to the state-of-the-art, while enabling training at scale.
    \item The \emph{LED} model achieves competitive performance compared to the state-of-the-art despite its simplicity.
    \item Pre-training embeddings on \emph{view} events and fine-tuning them using a projection matrix is an effective way to transfer knowledge to the \emph{click} prediction task.
\end{itemize}

Section~\ref{subsec:exp-setup} in Appendix details the parameters and experiment setup.

\subsection{Datasets}
\label{sec:datasets}
We use two datasets for the offline evaluation of our system.  The public dataset \textbf{ML20M} provides a benchmark that anchors our work within the existing literature.  The large-scale dataset \textbf{Production} demonstrates the ability of our approach to scale. Table~\ref{table:dataset-statistics} reports statistics about each dataset.

\begin{itemize}
    \item \textbf{ML20M}: A user-movie ratings dataset commonly used as a benchmark in the recommendation literature. We use the same preprocessing as \cite{variational-liang-2018} by keeping only ratings of four and higher and users with at least five ratings.
    \item \textbf{Production}: We collect user interactions with products on merchant websites and banners on publisher websites using the production system already in place for a period of three months.
 \end{itemize}

\begin{table}[h]
\centering
\caption{Dataset statistics. Density refers to the density of the item-item matrix.}
\label{table:dataset-statistics}
\begin{tabularx}{.46\textwidth}{|X|r|r|r|d{3.2}|}
  \hline
   & users &  items & events & \multicolumn{1}{|c|}{density \%}\\
  \hline
  ML20M   & 136K  & 20K & 10M  & 2.47  \\
  Production   & 587M  & 5M & 19B  & 0.076  \\
      \hline
\end{tabularx}
\end{table}

\subsection{Sampling loss}
\label{subsec-sampling-loss}
In this experiment, we compare the different losses of the \textit{LED} model on the ML20M dataset. Unless specified otherwise, we use $ N = 1000 $ negatives per user, shared across target positives. For all models, initialization is random, and \emph{Full} fine-tuning is enabled. For the \textit{Mult-VAE}, we use our scalable implementation using embedding lookups instead of one-hot vectors and observe the same results as those reported in the paper. We also include the performance of the \emph{Mult-VAE} trained with the approximated softmax (Mult-CSS). All losses except the BPR use the same items in both the input and target. For the BPR, we shuffle and randomly split the user timeline into input ($ 80\% $) and target ($ 20\% $). For the sake of comparison, we also show the results reported for other methods such as EASE~\cite{ease-vae},  WMF~\cite{wmf}, and SLIM~\cite{slim}.

\begin{table}[h]
\centering
\caption{Comparison of VAE and LED models with and without sampling on ML20M. Percentages measure relative difference with the \emph{Mult-VAE}. Only lines marked with a $\dagger$ are scalable. The LED model trained with BPR achieves results close to the Mult-VAE while enabling training at scale.}
\begin{tabularx}{.49\textwidth}{|X|X|l|l|}
  \hline
    \multicolumn{4}{|c|}{ML20M dataset} \\
  \hline
    Model & Loss & Recall@20 & Recall@50 \\
    \hline
  VAE~\cite{variational-liang-2018} & Mult & \textbf{0.396} & \textbf{0.537}   \\
  VAE & Mult-CSS $^\dagger$ &  0.382 (-3.54\%) & 0.523 (-2.61\%) \\
  DAE~\cite{variational-liang-2018} & Mult & 0.387 (-2.27\%)  & 0.524 (-2.42\%) \\
  \hline
  LED & Mult & 0.379 (-4.29\%) & 0.517 (-3.72\%) \\
  LED & Mult-CSS $^\dagger$ & 0.368 (-7.07\%) & 0.506 (-5.77\%) \\
  LED & BPR $^\dagger$ & \textbf{0.375} (-5.30\%) & \textbf{0.516}  (-3.91\%) \\
  LED & NS $^\dagger$ & 0.375 (-5.30\%) & 0.514  (-4.28\%) \\
  \hline
  EASE~\cite{ease-vae} & & 0.391 (-1.26\%) & 0.521  (-2.98\%) \\
  WMF~\cite{wmf} & & 0.360 (-9.09\%) & 0.498 (-7.26\%)\\
  SLIM~\cite{slim} & &0.370 (-6.57\%) & 0.495  (-7.82\%) \\
    \hline
    \end{tabularx}
\label{table:results-sampling-loss-ml20m}
\end{table}

Table~\ref{table:results-sampling-loss-ml20m} reports the results. From this experiment, we conclude that (1) training models using a sampling-based loss (\emph{Multi-CSS}\cite{pmlr-v54-botev17a}, \emph{BPR}\cite{bpr} or \emph{NS}\cite{NIPS2013_5021}) degrades the performance in a minimal way while enabling training at scale (2) the \emph{LED} model reaches close to state-of-the-art performance while bringing simplicity in the design. Similar results were obtained on the Production dataset. We select the BPR loss for further experiments on the Production dataset.

\subsection{Fine-tuning the embeddings}
\label{sec:fine-tuning-results}
Section \ref{subsec:fine-tuning-svd-embeddings} discusses two ways of fine-tuning pre-trained SVD embeddings. We start by artificially evaluating their efficiency on the ML20M dataset. Since this dataset contains only one type of events and is small enough for a model to perform multiple epochs on it, we emulate the large-scale scenario with the following procedure. We first train SVD embeddings on the full training set. Then, we train a \emph{LED} with multinomial likelihood on a small fraction of the training set and evaluate the effects of initialization and fine-tuning methods.

\begin{figure}[h!]
    \centering
    \includegraphics[width=0.4\textwidth]{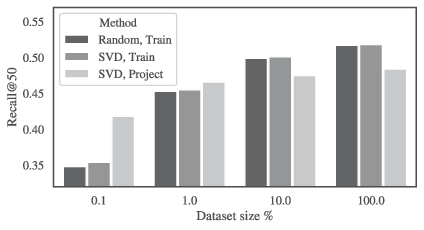}
    \caption{Recall@50 of \emph{LED} for different initialization and fine-tuning methods on ML20M. 
    The SVD embeddings are always pre-trained on the full training set. The \emph{project} method outperforms both random initialization and classical fine-tuning for models trained on small fractions of the dataset.}
    \label{fig:fine-tuning}
\end{figure}

Figure ~\ref{fig:fine-tuning} shows that the \emph{Project} method outperforms both random initialization and classical fine-tuning for models trained on $ 0.1 \% $ and $ 1 \% $ of the dataset. However, when the dataset size increases, training the embeddings yields better performance. The \emph{project} method still achieves decent results, which is impressive considering that it learns a $ d \times d $ matrix instead of the full embedding matrix $ V \in \mathbb{R}^{I \times d} $.

We go on to the experiments on the Production dataset for whose scale the pre-training method was designed. Table ~\ref{table:results-fine-tuning-products} summarizes the results. First, we find that on this dataset, the \emph{LED} model performs better than the \emph{VAE}. When training the embeddings (using either random or SVD initialization), we observe that the \emph{VAE} is more prone to over-fitting compared to the \emph{LED} model. The \emph{Project} fine-tuning method proves to be effective, yielding the best click-ranks for both models.

\begin{table}[h]
\centering
\caption{Impact of pre-training and fine-tuning methods on the \textbf{Production} dataset. Despite its simplicity, LED outperforms the VAE. Pre-training is particularly effective, yielding better results than random initialization.}
\label{table:results-fine-tuning-products}
\begin{tabularx}{.49\textwidth}{|X|X|X|l|l|}
  \hline
    \multicolumn{5}{|c|}{Production dataset} \\
  \hline
    Model & Init & Tuning  & R@20 & ClickRank \\
  \hline
    VAE & Random & Train & 0.078 & 0.471 \\
    VAE & SVD & Train & 0.083 & 0.457 \\
    VAE & SVD & Proj & \textbf{0.091} &\textbf{0.454} \\
  \hline
    LED & Random & Train & 0.099 & 0.468 \\
    LED & SVD & Train & \textbf{0.109} & 0.454 \\
    LED & SVD & Proj & 0.104 & \textbf{0.450} \\
  \hline
    \end{tabularx}
\end{table}

\section{System architecture}
\label{sec:system-architecture}

We propose an architecture that incorporates the methods described in the previous section suited for a real-world, large-scale deployment. We address scale and robustness through modularity by splitting the system into components that can be tested and improved independently.


At the highest level, our system is made of two components. First, a candidate selection step retrieves a list of candidate items from different algorithms (such as LED). Second, a ranking model predicts which of the candidates have the highest probability of being clicked or purchased and builds the final banner. The ranking model uses a logistic regression trained to predict clicks or sales.

Splitting the recommendation task into these two independent tasks is typical~\cite{amazon-deep-2019} but also known to be sub-optimal~\cite{off-policy-reco-www-2020}.  End-to-end training of two-stage systems remains impractical given our requirements (Section~\ref{subsec:requirements}). This approach, on the other hand, offers modularity and allows us to "fuse" different recommendation signals in the same system. Our online experiments show that this feature is useful in practice (Section~\ref{sec:online-results}). 

\subsection{Offline Pipeline}
\label{subsec:offline-pipline}

\begin{figure}[h!]
\centering
\includegraphics[width=0.43\textwidth]{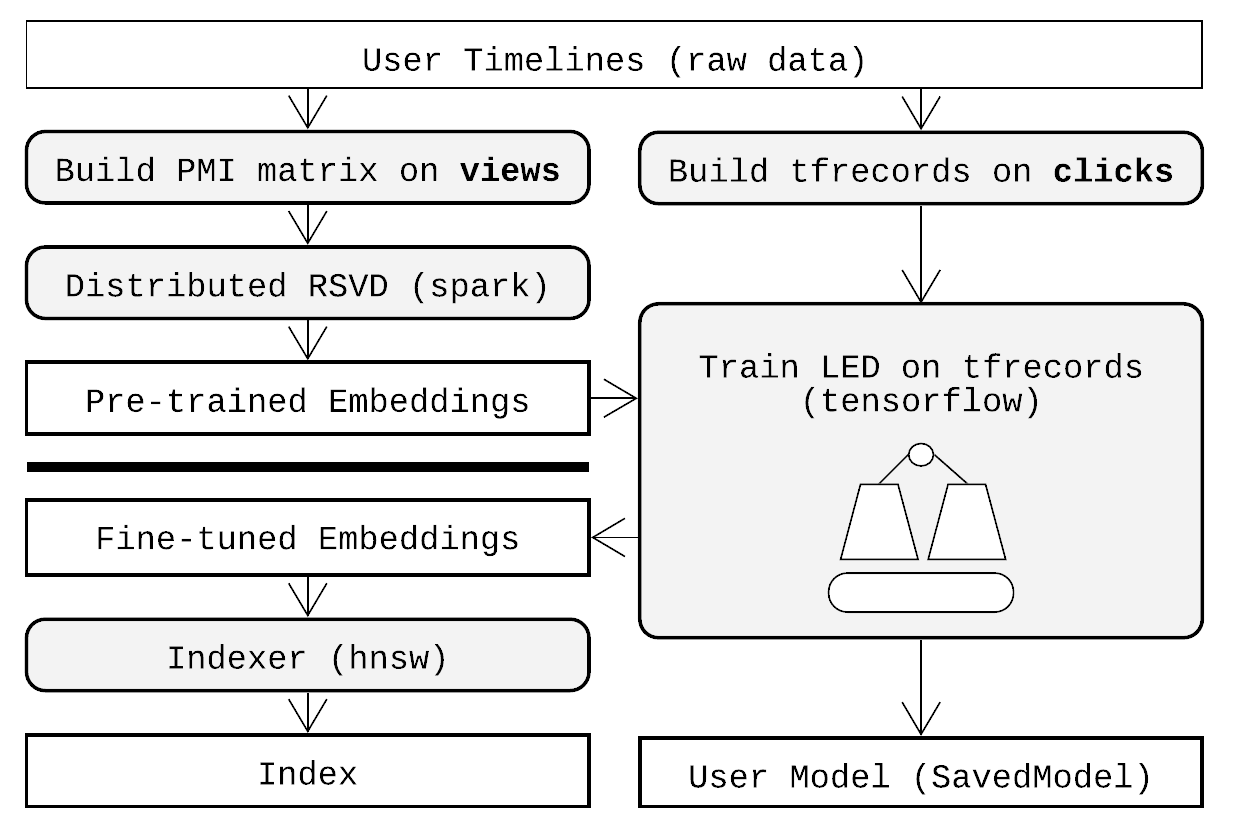}
\caption{LED: Overview of the offline pipeline to train the retrieval system}
\label{fig:reference-system}
\end{figure}

An offline component (Figure~\ref{fig:reference-system}) computes product embeddings from user timelines using RSVD which are then fine-tuned by training the LED model. Embeddings are further indexed into an appropriate data structure for fast retrieval (Maximum Inner Product Search). 

The data and models are split by country. The PMI, RSVD and dataset creation are distributed using spark, while the LED training is run on a single machine with Tensorflow. The model is exported as a Tensorflow Protocol Buffer file.


Saving computational graphs as Protocol Buffer files makes the pipeline generic as it can support any other model. Thanks to XLA: Ahead of Time Compilation (XLA AoT)~\footnote{\url{https://www.tensorflow.org/xla/tfcompile}}, these files can be further optimized for inference. XLA AoT compiles the model graph into machine instructions for various architectures through intermediate LLVM representation~\footnote{\url{http://llvm.org/}}. This yields significant performance improvement compared to a hand-coded implementation (x5 in our setting).

\subsection{Real-time retrieval at scale}
\label{subsec:online-knn}

An online JVM service loads both the product embeddings' indices and the user model in memory for several countries and starts processing requests via gRPC~\footnote{\url{https://grpc.io/}}. For each request, the online service receives the user's history and performs the following steps:
  \begin{enumerate}
    \item Fetch relevant product embeddings from hnsw.
    \item Pass embeddings and other features to LED and compute a user embedding using TensorFlow Java bindings ~\footnote{\url{https://www.tensorflow.org/install/lang_java}}
    \item Run approximate KNN search (hnswlib) with the user embedding as a query
    \item Return the approximate nearest neighbors as recommendations
\end{enumerate}


This architecture leverages offline computation for the product embeddings and model training while using fresh user data to compute recommendations.  Regarding the nearest-neighbor search, the scale of our problem precludes the use of exact search. Approximate nearest neighbor is an attractive alternative and has been extensively studied. We built on top of existing benchmarks~\footnote{\url{http://ann-benchmarks.com}} and chose hnswlib~\cite{hnswlib-2018}. 



This design has several advantages. First, it enables to leverage the full power of representation learning throughout the whole pipeline. Indeed, product and user representations remain accessible in their full vector state until the final recommendation is done. Second, the memory footprint is independent of the number of users and only scales linearly with the number of products. Third, it fails gracefully in case of user cold start: users with no history get recommendations by skipping the user embedding computing step and doing a nearest neighbour search with a null vector which is equivalent to retrieving the products with the highest popularity bias for the merchant. Section~\ref{subsec:compute-resources} shows that minimal hardware is required to serve billions of users with this architecture.

\subsection{Computation resources}
\label{subsec:compute-resources}

In production, the offline pipeline is made of data pre-processing, RSVD, model training, and hnsw indexing. It runs on our internal Hadoop cluster independently for each country.
The RSVD and indexing jobs are parallelized using Spark on a few dozen executors with 16 CPU and 30-GB RAM, depending on the number of items in a country's catalog. Each Tensorflow model is trained on a single machine with 48 CPU and 48-GB RAM.
The whole pipeline, including training on the full dataset, runs in a few hours and is scheduled every day to incorporate new data. The online service runs as a pool of instances within an Apache Mesos cluster. Each instance has 4 CPU, 30~GB RAM, and serves all countries. 


The system must handle billions of requests per day. Hence, we monitor the long tail of inference time distribution to minimize the number of servers needed and production incidents. Some useful numbers describing the system scalability are listed in Table~\ref{table:system-numbers}. We report latency including request deserialization and network overheads. LED provides an end-to-end recommendation in just 2~ms on a single server, i.e. $30\times$ faster than the best system known so far~\cite{pixie-pinterest-www-2018}.

\begin{table}[h]
\centering
\caption{Real-time computing performance of our system with the LED model}
\begin{tabularx}{.46\textwidth}{|X|l|}
  \hline
    Max Queries Per Second (QPS) per instance & 3200 \\
  \hline
    Latency @ 50th pct & 500$\mu$s \\
    Latency @ 99th pct & 2ms \\
  \hline
    Latency of user embedding computation @ 50th pct & 30$\mu$s \\
    Latency of user embedding computation @ 99th pct & 65$\mu$s \\
  \hline
    Latency of KNN search @ 50th pct & 160$\mu$s \\
    Latency of KNN search @ 99th pct & 450$\mu$s \\
  \hline
    Instances used in production & 200 \\
  \hline
    Recommendations served per day & 4B \\
  \hline
    \end{tabularx}
\label{table:system-numbers}
\end{table} 


\section{Online evaluation}
\label{sec:online-results}

We present here the results of live testing of our system on real traffic. This experiment serves two purposes: first, it demonstrates the ability of our system to operate at the scale specified in our requirements; second, it proves that the performance gain observed in offline experiments translates into actual business uplift once tested with real users.

We use a classical A/B testing approach where subsets of randomly selected users are exposed to various algorithms. In these experiments, the ranking model is trained continuously several times a day and is the same for all populations. In particular, no feature is added to the ranking model that would benefit a particular method or population.

\subsection{A/B test setup}
\label{sec:ab-test-setup}
We use three algorithms for the A/B test:
\begin{itemize}
    \item \textbf{Global Best Of (GBO)}: This algorithm recommends the most popular items of the merchant to each user, independently from their history. This algorithm is simple and provides a reference for a non-personalized recommendation.
    
    \item \textbf{Clustering Best Of (CBO)}: This algorithm corresponds to the reference algorithm that was running in production before the project started. The algorithm leverages user clustering and computes a top-k for sets of users sharing similar interests in items. CBO leverages cross-merchant information and was developed and optimized over several years, thus representing a strong baseline for our work.
    
    \item \textbf{LED}: This is the algorithm chosen after offline experimentation as described in Section~\ref{sec:offline-evaluation}, i.e. with RSVD initialization and fine-tuning by Projection.
\end{itemize}

Before the A/B test, all users are exposed to the baseline algorithms, i.e. GBO + CBO. During the A/B test, we split users at random into three groups: A, B, and C. Users in group A are exposed to GBO products only. Users in group B are  exposed to GBO + CBO products. Finally, users in group C are exposed to products coming from GBO + CBO + LED. We use sets of algorithms instead of individual algorithms as it lets the ranking algorithm learn which algorithm works best for each population dynamically. This approach is typical in an industrial setting where a new algorithm (LED) is evaluated against an existing baseline running in production (GBO + CBO).

We consider three evaluation metrics: clicks on banner products, sales associated with a display, and total sales amount (order value). Sales are sparser than clicks but better capture the interest of the user. The large volume of our test allows us to draw conclusions on sales well out of noise.

To better capture the interest of the user, we used landed clicks, i.e. clicks that actually drove the user to a product page on the merchant website. This definition of click is sparser but less noisy and better represents true user interest, since unintentional clicks and failed landing pages are discarded. Sales correspond to actual sales on the merchants' websites.

The test lasts two months and collects $2$~billion displays. The displayed products contain a mixture of goods, services, travel and classified ads and covers thousands of merchants across Europe, America and Asia.

\subsection{A/B test results}
\label{ab-test-results}
\begin{figure}[h]
    \centering
    \includegraphics[width=0.49\textwidth]{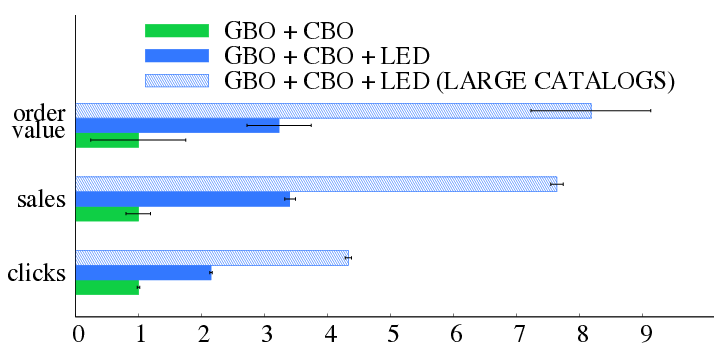}
    \caption{A/B test results: uplift of GBO + CBO and GBO + CBO + LED versus GBO.  The uplift of GBO + CBO is scaled to 1. Error bars represent confidence intervals at 95\%.}
    \label{fig:ab-test-results}
\end{figure}

\begin{figure}[h]
    \centering
    \includegraphics[width=0.49\textwidth]{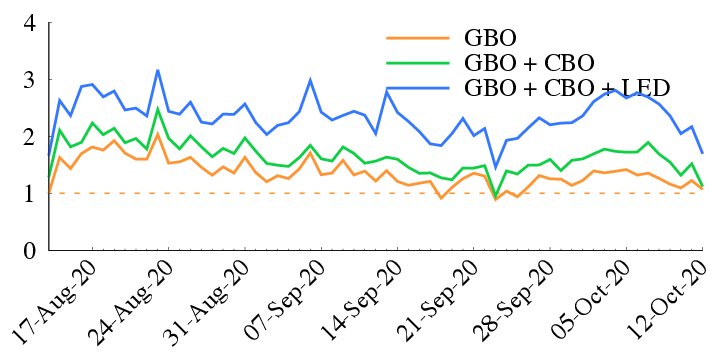}
    \caption{Number of sales per day for each population of the A/B test. The test lasts two months. Y-axis scaled to 1 for the first day of GBO.}
    \label{fig:ab-test-results-daily}
\end{figure}

 Figure~\ref{fig:ab-test-results} shows the uplift of GBO + CBO + LED over GBO normalized by the uplift of CBO over GBO. Error bars represent $95\%$ confidence intervals. We observe a significant increase in clicks, sales and order value (sales amount). Specifically, LED brings twice more landed clicks and 3.5 times more sales than GPO + CBO did over GBO. In other words, representation learning increased multipled times the added value of recommendation in our system. The uplift is observed continuously over the A/B test period (Figure~\ref{fig:ab-test-results-daily}).
 
 We also analyze the uplift for a specific set of merchants that possess large and diverse catalogs. Intuitively, we expect a larger uplift of our method for these merchants for which the room for recommendation is larger. In practice, merchants belong to this group if they have more than $10K$ items and more than $10$ different categories in their catalog. Categories are associated with each product based on the Google taxonomy using an independent method. With this definition, we indeed observe a higher uplift for these merchants. The uplift in clicks increases four-fold while the uplift in sales increases eight-fold.

Table~\ref{table:source-share-of-voice} shows the share of products displayed for each algorithm in each population. We remind that the ranking algorithm does not favor a particular algorithm (GBO, CBO, or LED) and only selects the best product to optimize clicks and sales. We observe that LED gains a significant portion of the displayed products ($68\%$) at the expense of GBO and CBO. This testifies of the strong interest of users for the products recommended by LED. GBO and CBO maintain a significant share of voice in situations where popularity is a more effective approach than personalization. The advantage of a modular and two-step architecture is obvious here.

\begin{figure}[h]
    \centering
    \includegraphics[width=0.49\textwidth]{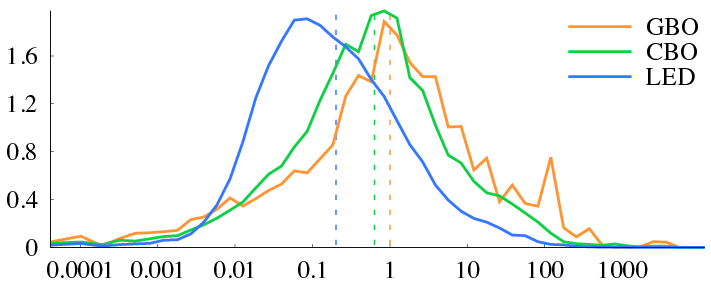}
    \caption{Distribution of product popularity per algorithm. x-axis: number of views per month on a log scale normalized to 1 for GBO. Average value in dotted line. Despite showing less popular products, LED generates more clicks and sales.}
    \label{fig:histogram-popularity}
\end{figure}

Figure~\ref{fig:histogram-popularity} shows the distribution of product popularity for each algorithm. Unsurprisingly, GBO products have the highest popularity overall. CBO products are 3x less popular on average, while LED returns products that are 10x less popular than GBO. LED recommendations venture deep into the long tail of the popularity distribution, while still achieving state-of-the-art performance as seen in Section~\ref{sec:ab-test-setup}. This result is particularly interesting since the popularity bias of LED is learned by the model and not tuned manually (Section~\ref{sec:models}). Showing less popular products is a valuable behavior from a user experience point of view, resulting in more product discovery for end users.

From these results, we conclude that LED generates recommendations that are significantly more interesting for the users and the merchants. The increase is significant both in clicks and in sales.

\begin{table}[h]
\centering
\caption{Share of each algorithm in the displayed products after ranking. The ranking model predicts clicks and sales independently from the product origin.}
\begin{tabularx}{.4\textwidth}{|X|r|r|r|}
  \hline
   A/B test population & GBO & CBO & LED \\
    \hline
  A & 100\%  & 0\% & 0\%  \\
  \hline
  B & 61\%  & 39\% & 0\%  \\
   \hline
  C  & 22\%  & 11\% & 68\% \\
    \hline
    \end{tabularx}
\label{table:source-share-of-voice}
\end{table}

\section{Conclusions and Future work}

Recommendation engines have become an essential part of online services. Systems operating at the scale of billions of users have been extensively described. Yet, new algorithms such as variational auto-encoders have raised the bar in terms of algorithmic performance but present significant limitations of scale, both for training and inference.

We propose a simple and efficient model (LED, for Lightweight Encoder-Decoder) reaching a new trade-off between complexity, scale and performance. By combining pre-training, sampled losses and amortized inference, LED brings a $30\times$ speed-up in latency while reaching the same performance as variational auto-encoders~\cite{variational-liang-2018} on standard recommendation metrics.

The model combines several key design choices such as pre-training, amortized inference, sampling-based losses, and fast nearest-neighbor search. We provided a detailed description of a system as deployed at Criteo and open source our code, making our work useful for practitioners wishing to deploy an efficient large-scale recommendation system in the real-world. 

Future work will examine the added value of side information in the architecture. The ability to inject more diversity in the recommendation is also a valuable area of research.


\bibliographystyle{ACM-Reference-Format}
\bibliography{main}

\newpage
\section{Reproducibility}

\subsection{Code}
We release the code used for our experiments under the Apache 2.0 License.~\footnote{https://github.com/criteo/deepr}. The ML20M experiments can be reproduced on a standard machine with Python 3.x and standard libraries.

\subsection{Experimental setup}
\label{subsec:exp-setup}

We describe the parameters used to obtain the results described in Sections~\ref{subsec-sampling-loss} and~\ref{sec:fine-tuning-results}.

\textbf{ML20M dataset} We use the same experimental setup as the \emph{Mult-VAE} paper~\cite{variational-liang-2018} and split the users into train / validation / test, which means one user does not appear simultaneously in train and validation / test. At test time, we randomly split timelines into input ($ 80\% $) and target ($ 20\% $) and evaluate the model's recommendations as a top-$k$ retrieval task by reporting recall at $ k=20 $ and $ k=50 $. 

For the \emph{VAE}, we use embeddings of dimension $ d = 600 $, hidden layers of dimension $ 200 $, $ \text{tanh} $ activation, and anneal the KL divergence from $ 0 $ to $ 0.2 $ during training. The embeddings' dimension is set to $ d = 600 $ for the \emph{LED} as well and we apply denoising with probability $ 0.5 $ to both models. We use the Adam~\cite{adam-iclr-2015} optimizer with learning rate $ 0.001 $, batch size $ 512 $, and select the best checkpoint using NDCG@100 on the validation users. All models are trained for 50k steps ($ \sim 200 $  epochs) with a checkpoint frequency of 230 steps (roughly every epoch). Unless specified otherwise, we use $ N = 1000 $ negatives sampled uniformly.

\textbf{Production dataset} We keep the last 7 days of each user activity for testing, and split the remaining days by user into $90\%$ for training and $10\%$ for validation. For each user in training and validation, we use the last 7 days as the target part and the first days as the input part. Because we only keep timelines with at least one click in the target period, the resulting pre-processed dataset is only a fraction of the original dataset. Although we are interested in the top-$k$ retrieval task, we are also focused on the quality of the top recommendations and report click-rank in addition to recall at $20$. Click-rank is defined as the normalized rank of a clicked item among other items in one banner sorted by score and typically goes from $ 0.5 $ (random system) to $ 0 $ (perfect system, clicked item has the highest score returned by the model).

The embedding size is an important hyper-parameter. We experimented with various values ranging from $10$ to $1000$ and found little performance improvement beyond $100$. We choose an embedding size of $100$ for both the \emph{LED} and \emph{VAE}. We use LazyAdam with learning rate $ 0.001 $ and batch size $ 512 $. We train models for 100k steps ($ \sim 30 $ epochs) with a checkpoint frequency of 2k. We also reuse the same embeddings for the input and output layers of the \emph{VAE}. Unless specified otherwise, the \emph{VAE} is trained with approximated multinomial likelihood (Mult-CSS), while the \emph{LED} is trained with BPR. Using denoising did not seem to increase performance. As banners usually contain a few items (typically between 2 and 10), we use the non-clicked items as negatives. 

\subsection{Significant parameters}

The sampling of negatives is a significant parameter to tune and should be carefully set as more negatives come with a cost in training time. The following experiment evaluates the impact of the number of negatives on performance. We train a \emph{LED} with BPR for a varying number of negatives and compare the resulting metrics with the same model trained with a multinomial likelihood. 

Unsurprisingly, increasing the number of negatives yields better performance. With $ N = 1000 $ negatives, the BPR is within $ 1\% $ of the metrics of the Multinomial Likelihood. However, even with low values ($ N = 10 $), the relative difference does not exceed $ 5\% $. We observe similar trends with other sampling-based losses or models.

\begin{figure}[h!]
    \centering
    \includegraphics[width=0.46\textwidth]{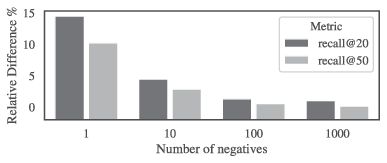}
    \caption{Relative performance drop of \emph{LED} trained with BPR instead of multinomial likelihood (smaller is better) on ML20M. With only 10 negatives, the drop is less than 5 \%.}
    \label{fig:num-negatives}
\end{figure}

In Section~\ref{subsec:LED}, the normalization factor $ \frac{1}{T} $ is less common than $ \frac{1}{\sqrt{T}} $. In practice, we obtained slightly better results with it. Intuitively, it controls the relative importance with the biases. On one hand, using $ \frac{1}{T} $ means that the bias will have more importance when the user's history is diverse, while short and long homogeneous timelines will be treated equally (consider a timeline made of $ T $ times the same item). On the other hand, using $ \frac{1}{\sqrt{T}} $, decreases the importance of the bias for long homogeneous timelines compared to short ones, while treating diverse timelines similarly to short but homogeneous ones.

In Section~\ref{subsec:matrix-fatorization-with-randomized-svd}, we use an implementation of distributed randomized SVD on Spark already open sourced by Criteo~\footnote{\url{https://github.com/criteo/Spark-RSVD}}. As in~\cite{levy2015improving}, we observe that smoothing the distribution of context products improves performance. We implement this by raising the probability $p(j)$ to a power strictly lower than $1$ (we choose $0.75$ as is commonly done in the literature).

\subsection{Qualitative evaluation}

\begin{figure}[h]
    \centering
    \includegraphics[width=0.4\textwidth]{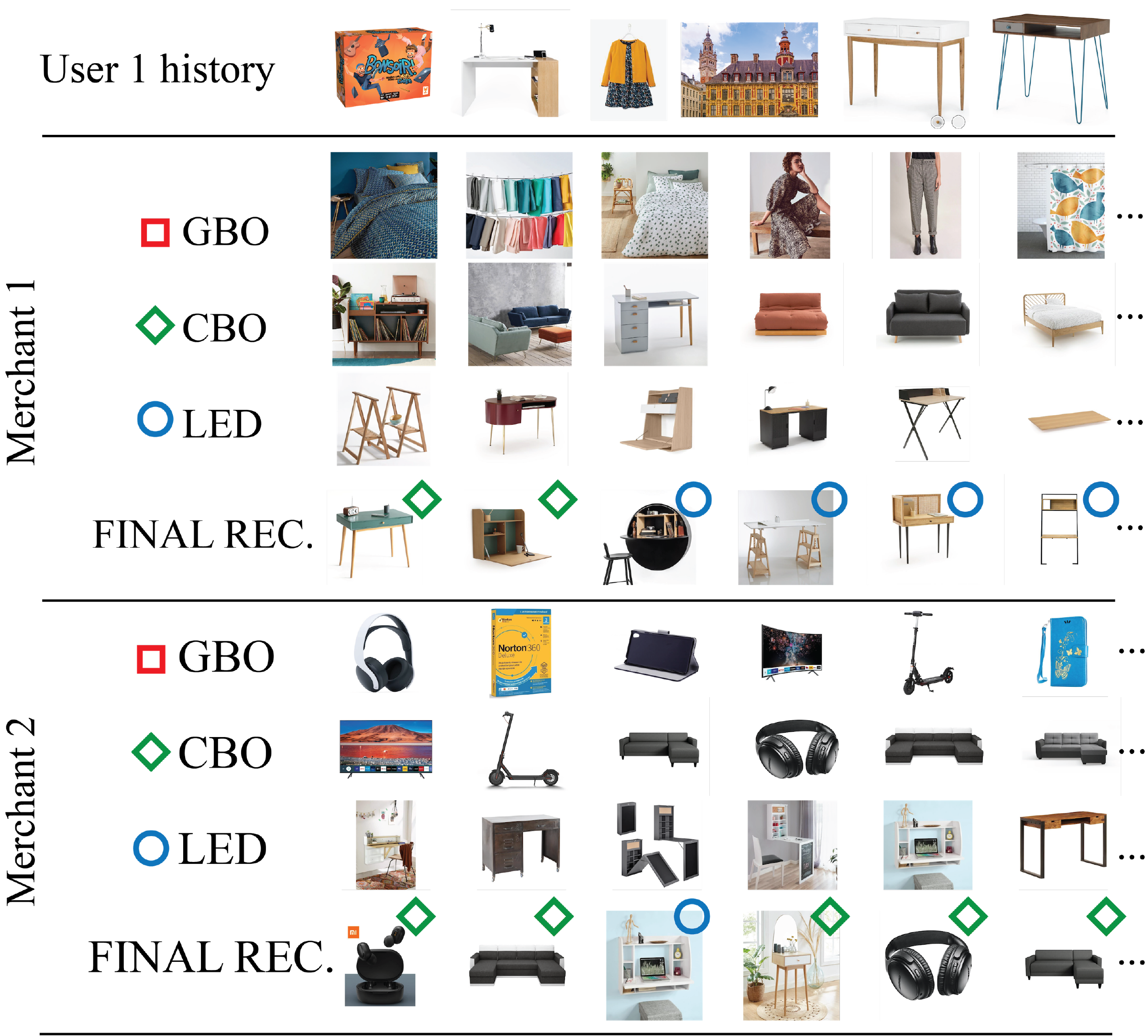}
    \includegraphics[width=0.4\textwidth]{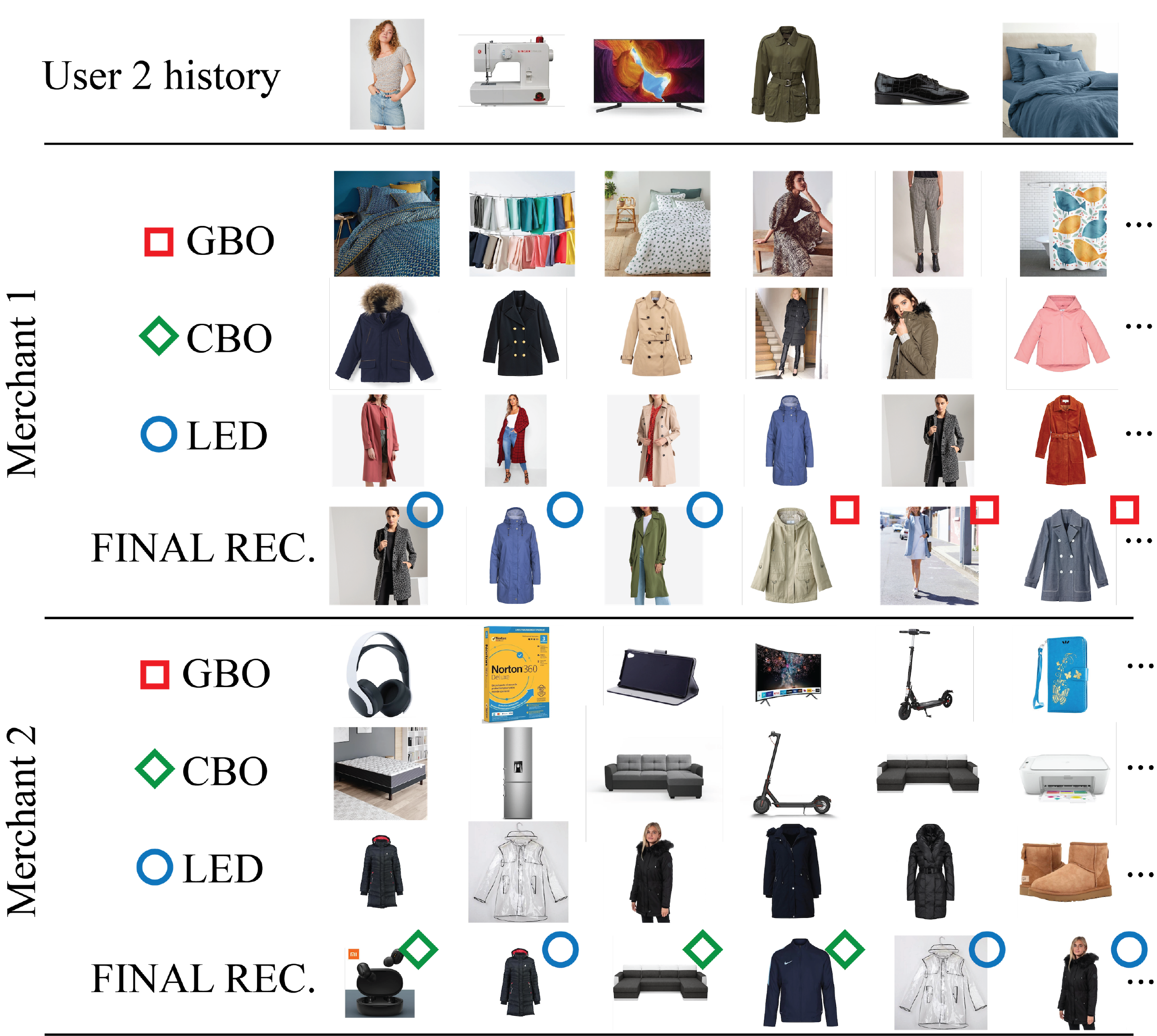}
    \caption{Sample recommendations for two users and two merchants. LED recommends more personalized products thanks to its fine-grain product-level user representation.}
    \label{fig:viewer-results}
\end{figure}

Qualitative evaluation helps understanding the behavior of the algorithm. We do this by selecting several real user histories at random and inspecting the algorithm output. Figure~\ref{fig:viewer-results} shows products recommended by GBO, CBO and LED for two merchants and two real users. Merchant 1 sells mostly furniture and clothes. Merchant 2 is a more general retailer selling a larger variety of items such as furniture, electronics, hardware, and software. User 1 is mostly interested in desks, traveling and children clothes; User 2 browsed women's clothes, sewing machine and TVs. For each user, we show six products sampled from their browsing history. The last row of each set shows the final recommendation output by the ranking model. A colored label identifies the algorithm which recommended each product. We extract several lessons from this analysis:

\begin{itemize}
    \item CBO manages to capture some of the interests of the user; by design, this algorithm is less personalized. On the other hand, LED leverages product-level representations and yields more personalized recommendations.
    \item LED is surprisingly robust and manages to extract useful recommendations even though some items from the user history are irrelevant to the merchant. This is particularly visible for User 1 and Merchant 1 where LED recommends desks despite travel items being present in the user history, and User 2 and Merchant 2 where LED extracts an interest in coats and shoes among diverse user interests.
    \item LED recommends items that are not necessarily similar to the items browsed by the user but also complementary (e.g. desk board or trestles for User 1 and Merchant 1).
    \item LED recommendations are fairly narrow and focused on a specific type of product. For instance, for User 2 and Merchant 1, recommendations for bed covers (which are available for this merchant as seen in GBO) could be considered as relevant. An interesting area for future work is to build multi-modal user representations.
\end{itemize}

\subsection{Relationship to other methods} 
In the aim of clarifying our algorithm as much as possible, we explain the differences it bears with similar algorithms existing in the literature.

When trained as an auto-encoder with multinomial likelihood, the \emph{LED} model is the lightest possible \emph{Mult-DAE}~\cite{variational-liang-2018}. It has no hidden layers and shares both \emph{encoder} and \emph{decoder} embeddings. The LED design is motivated by the unique scale at which we operate, constraining us to reduce compute and memory footprint to the minimum. We note that both the \emph{Mult-DAE} and \emph{Mult-VAE}~\cite{variational-liang-2018} satisfy the \emph{amortized inference} and \emph{fast nearest neighbor search} guidelines since the model parameters are shared by all users and the weights of the last softmax layer can be interpreted as item embeddings.

The LED model bears a lot of similarities with other latent factors models. If we omit the normalization factor, the biases, and use a gaussian likelihood as objective, it is equivalent to minimizing $ \Vert X - X \cdot V \cdot V^T \Vert^2_F $, where $ X \in \mathbb{N}^{U \times I} $ is the user-item interaction matrix, $ V \in \mathbb{R}^{I \times d} $ the item embeddings matrix, and $ \Vert \cdot \Vert_F $ the Frobenius norm. This formulation is closely related to standard SVD of the user-item matrix $ X $, since its closed-form solution is the matrix formed of the top-$d$ right singular vectors. It is also similar to the SLIM~\cite{slim} objective, with the sparsity constraint replaced with a low-rank constraint. This does not come as a surprise since the \emph{VAE} framework is known to be a generalization of latent factors models~\cite{variational-liang-2018}.

Previous attempts to simplify matrix factorization methods include NSVD~\cite{paterek}, which also represents users using their history. In particular, it uses a Gaussian likelihood to reconstruct recommendation scores as $ s(u, i) = b_u + b_i + \left\langle \sum_{t=1}^T \overrightarrow{v_{u_t}}, \overrightarrow{v_i} \right\rangle $, where $ b_u $ (resp. $ b_i $) are user (resp. item) biases. The \emph{LED} differs mostly by the absence of user biases, the normalization factor, as well as the training procedure.

As an item-based model, our approach can also be seen as a simple neighborhood-based method~\cite{neighbors}. Building on the related NSVD ~\cite{paterek} method, more powerful variants~\cite{koren} intersecting matrix factorization with neighborhood-based techniques have been explored.

If we apply a softmax to the scores to get a distribution over items, this model is a one-hidden-layer neural network, whose input is a normalized one-hot encoding of the user's history and output a distribution over items. If we substitute items with words, we recognize a log-linear model, similar to Word2Vec~\cite{NIPS2013_5021}, where context and target parameters are shared and inputs normalized.

In Section~\ref{subsec:fine-tuning-svd-embeddings}, $ P^T \cdot P $ is the Gram matrix associated with the kernel $ K(x, y) = <Px, Py> $. Previous work~\cite{kernel-mf} proposed a regularized matrix factorization method using kernels and derived an online update rule to solve the new user/item problem. Instead of updating the user/item matrices, our method focuses on updating the kernel itself to adapt to a new type of feedback.


\end{document}